\documentclass[aps,prl,twocolumn,groupedaddress]{revtex4-1}
\usepackage{bm}
\usepackage[pdftex]{graphicx}
\usepackage{here}
\usepackage{amsmath,amsfonts,amssymb}
\usepackage{hyperref}
\usepackage{comment}
\usepackage{xcolor}
\usepackage{float}

\begin{document}
% \pagewiselinenumbers
\title{Bulk-edge correspondence for nonlinear eigenvalue problems
		}

\author{Takuma Isobe$^1$}
\author{Tsuneya Yoshida$^{2}$}
\author{Yasuhiro Hatsugai$^{1,3}$}
\affiliation{
$^1$Graduate School of Pure and Applied Sciences, University of Tsukuba, 
Tsukuba, Ibaraki 305-8571, Japan\\
$^2$Department of Physics, Kyoto University, Kyoto 606-8502, Japan,
$^3$Department of Physics,
University of Tsukuba, Tsukuba, Ibaraki 305-8571, Japan
}

\date{\today}

%%%%%%%%%%%%%%%%%%%%%%%%%%%%%%%%%%%%%%%%%%%%%%%%%%%%%%%%%%%%%%%%%%%%%%%%%
%%%%%%%%%%%%%%%%          abstract                          %%%%%%%%%%%%%
%%%%%%%%%%%%%%%%%%%%%%%%%%%%%%%%%%%%%%%%%%%%%%%%%%%%%%%%%%%%%%%%%%%%%%%%%
\begin{abstract}
Although topological phenomena attract growing interest not only in linear systems but also in nonlinear systems, the bulk-edge correspondence under the nonlinearity of eigenvalues has not been established so far.
We address this issue by introducing auxiliary eigenvalues.
We reveal that the topological edge states of auxiliary eigenstates are topologically inherited as physical edge states when the nonlinearity is weak but finite (i.e., auxiliary eigenvalues are monotonic as for the physical one). 
This result leads to the bulk-edge correspondence with the nonlinearity of eigenvalues.
\end{abstract}

\maketitle
%%%%%%%%%%%%%%%%%%%%%%%%%%%%%%%%%%%%%%%%%%%%%%%%%%%%%%%%%%%%%%%%%%%%%%%%%
%%%%%%%%%%%%%%%%          introduction                      %%%%%%%%%%%%%
%%%%%%%%%%%%%%%%%%%%%%%%%%%%%%%%%%%%%%%%%%%%%%%%%%%%%%%%%%%%%%%%%%%%%%%%%

%%%%%%%%%%%%%%%%%%%%%%%%%%%%%%%%%%%
%%%%%%%  paragraph1  %%%%%%%%%%%%%%
%%%%%%%%%%%%%%%%%%%%%%%%%%%%%%%%%%%
\textit{Introduction.}---The topological phase of matter has attracted considerable interest due to its exotic nature. In particular, topological band theory has played a crucial role in unveiling various types of topological phases by combining the principles of band theory and topology~\cite{C.L.Kane_E.J.Mele_PRL.2005,C.L.Kane_E.J.Mele_PRL.2005_Z2,L.Fu_C.L.Kane_PRL.2007,M.Z.Hasan_C.L.Kane_RevModPhys.2010,X.L.Qi_S.C.Zhang_RevModPhys.2011,Y.Ando_JPSJ_2013,B.A.Bernevevig_T.L.Huglhes_S.C.Zhang_Science_2006,M.Knig_Science_2007,L.Fu_C.L.Kane_PRB.2007,L.Fu_C.L.Kane_PRB.2006,D.J.Thouless_PRB.1983,Schnyder_PRB.2008,A.Y.Kitaev_AIP_Conf_2009,S.Ryu_A.P.Schnyder_A.Furusaki_New.J.Phys_2010,X.L.Qi_T.L.Hughes_S.C.Zhang_PRB_2008,A.M.Essen_J.E.Moore_D.Vanderbilt_PRL_2009,S.Murakami_IOP_2007,W.Xiang_PRB_2011,Yang_PRB_2011,A.Birkov_PRL_2011,Xu_PRL_2011,Kurebayashi_JPSJ_2014,N.Armitage_RevModPhys_2018,Koshino_PRB_2016}.
One of the most intriguing phenomena of topological phases is the bulk-edge correspondence (BEC), which indicates the edge states induced by the bulk-topology~\cite{Hatsugai_BEC_PRL(1993),Hatsugai_BEC_PRB(1993)}.
Such topological edge states emerge regardless of other details of systems and are sources of anomalous behaviors. For instance, the robust edge states in integer quantum Hall systems result in quantized Hall conductance with extreme accuracy~\cite{Klitzing_IQHE_PRL(1980),Laughlin_IQHE_PRB(1981),TKNN_IQHE_PRL(1982),Kohmoto_IQHE_AoP(1985),Haldane_AIQHE_PRL(1988)}.
Notably, in these years, the notion of topological states and BEC has expanded to encompass a broad range of systems,~\cite{Raghu_PhC_PRL(2008),Raghu_PhC_PRA(2008),MIT_PhChIns_PRL(2008),Lu_TopPhot_Nat(2014),Hu_TopPhot_PRL(2015),Takahashi_Optica(2017),Takahashi_JPSJ(2018),Ozawa_TopPhot_RMP19,OtaIwamoto_NatPhoto(2020),Moritake_NanoPh(2021),Kariyado_MechGraph_Nat(2015),Yang_TopAco_PRL(2015),Huber_TopMech_Nat(2016),Susstrunk_MechClass_PNAS(2016),Tomoda_AIP(2017),Kawaguchi_Nature(2017),Takahashi_Mech_PRB(2019),Liu_TopPhon_AFM(2020),Lee_TopCir_Nat(2018),Yoshida_Difus_Nat(2021),Makino_Difus_PRE(2022),Hu_ObsDifs_AM(2022),Knebel_GameTheor_PRL(2020),Yoshida_GameTheor_PRE(2021)}.
  including  interdisciplinary systems, such as meteorological systems~\cite{Delplace_TopoMeteo_science(2017)}.

Along with the above progress, generalizing topological band theory has also elucidated exotic phenomena. 
For instance, while topological band theory originally developed for systems with Hermitian eigenvalue problems, generalizing it to non-Hermitian systems has discovered the emergence of exceptional points~\cite{Zhen_ERing_nature(2015),Kozii_nH_arXiv(2017),Shen_NHTopBand_PRL(2018),Yoshida_NHhevferm_PRB(2018),Zyuzin_NHWyle_PRB(2018),Takata_pSSH_PRL(2018),Budich_SPERs_PRB(2019),Yoshida_SPERs_PRB(2019),Yoshida_PTEP(2020),Yoshida_SPERsMech_PRB(2019),Okugawa_SPERs_PRB(2019),Zhou_SPERs_Optica(2019),Mandal_HighEP_PRL(2021),Delplace_EP3_PRL(2021),IYH_PRB(2021),IYH_Nanoph(2023)} and skin effects~\cite{T.E.Lee_PRA_2016,Shunyu_PRL(2018)_SkinEffect,Flore_skin_PRL(2018),Yoshida_MSkinPRR20,Yokomizo_nbloch_PRL(2019),Borgina-Jan_PRL(2020),Okuma_skin_PRL(2020),CFang_skin_PRL(2020)} which do not have Hermitian counterparts.

In this respect, generalizing the topological band theory to nonlinear systems also induces novel phenomena. 
Indeed, the interplay between topology and nonlinearity of the eigenvectors has recently been discussed in Refs.~\cite{Bloch_TopoSoliton_PRL(2019),Ezawa_NLTopo_PRB(2022),Ezawa_NLTopo_PRB(2022),Ezawa_NLTopo_JPSJ(2022),Sone_TopoSync_PRR(2022),Bloch_TopoSoliton_Nature(2022),Jezequel-Delplace_PRB_(2022),Sato-Fukui_TopoTodaLatt_JPSJ(2023),Sone_NLTopo_arxiv(2023)}, elucidating topological synchronization induced by interplay between the nonlinearity and the topology~\cite{Sone_TopoSync_PRR(2022)}.
Despite the above extensive efforts, the interplay between the topology and nonlinearity of eigenvalues, another type of nonlinearity, has not been discussed so far. 
In particular, BEC, which plays a central role, has not been established in nonlinear systems of eigenvalues. 
The significance of this issue is further enhanced by the existence of relevant systems; some of photonic crystals and mechanical metamaterials are described by the nonlinear eigenvalue problem~\cite{Kuzmiak_Metal-PhC_PRB(1994),Huang_Mass-in-Mass_IJES(2009)}.

In this letter, we establish the BEC for nonlinear systems of eigenvalues. 
Our strategy is based on an auxiliary eigenvalue.
Introducing the auxiliary eigenvalue allows us to analyze the auxiliary edge states induced by the bulk-topology. Among them, focusing on the physical edge states leads us to the BEC under the nonlinearity of eigenvalues.
We demonstrate the emergence of edge states due to nonlinear BEC for two-dimensional insulators and three-dimensional semimetals.
Our approach of the auxiliary eigenvalue is considered to be versatile; it can be extended to systems in other symmetry classes and dimensions.

%%%%%%%%%%%%%%%%%%%%%%%%%%%%%%%%%%%%%%%%%%%%%%%%%%%%%%%%%%%%%%%%%%%%%%%%%
%%%%%%%%%%%%%%%%          GEVP with Hermitian matrices      %%%%%%%%%%%%% %%%%%%%%%%%%%%%%%%%%%%%%%%%%%%%%%%%%%%%%%%%%%%%%%%%%%%%%%%%%%%%%%%%%%%%%% 　\\ %%%%%%%%%%%%%%%%%%%%%%%%%%%%%%%%%%% %%%%%%%  paragraph1  %%%%%%%%%%%%%%
%%%%%%%%%%%%%%%%%%%%%%%%%%%%%%%%%%%
%%%%%%%   figure 3   %%%%%%%%%%%%%%
%%%%%%%%%%%%%%%%%%%%%%%%%%%%%%%%%%%
\begin{figure}[t]
   \includegraphics[width=8cm]{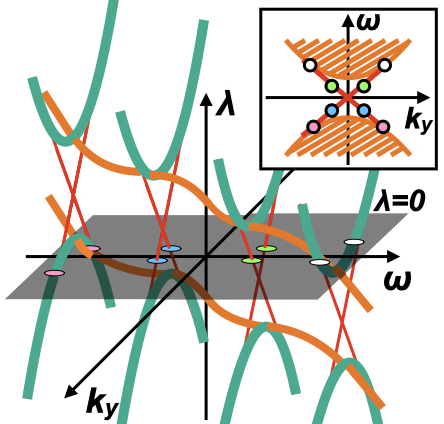}
 \caption{
Schematic picture of the auxiliary bands $\lambda(k,\omega)$ for two-dimensional cylindrical system. 
The system is periodic in $y$-direction and with boundaries for $x$-direction.
 Bulk states are illustrated in green, while edge states are represented in red for each omega. The eigenstates that cross the $\lambda=0$ plane are of particular interest in the context of physics.
Physical bands of $\omega(k)$ for each $k$ is shown in the inset. The gapless edge states of the auxiliary bands $\lambda$ are inherited to the physical nonlinear bands of $\omega$.
Namely, red, blue, green, and white dots on the gray plane of $\lambda=0$  respectively corresponds to the physical edge states. For example, they can be dots in the inset.
}
\label{label:Fig_model}
\end{figure}

%%%%%%%%%%%%%%%%%%%%%%%%%%%%%%%%%%%
\textit{Auxiliary eigenvalue and nonlinear bulk-edge correspondence.}---Here, we provides our strategy to discuss the nonlinear BEC, i.e., the BEC of the nonlinear eigenvalue problems [see Eq.~(1)].
We introduce the auxiliary eigenvalues and discuss the BEC between the bulk topology of the auxiliary bands and the physical edge states.

%%%%%%%%%%%%%%%%%%%%%%%%%%%%%%%%%%%
%%%%%%%  paragraph2  %%%%%%%%%%%%%%
%%%%%%%%%%%%%%%%%%%%%%%%%%%%%%%%%%%
We consider the following nonlinear equation.
\begin{equation}
H(\omega,\bm{k})\psi = \omega S(\omega,\bm{k})\psi,
\end{equation}
which is a nonlinear eigenvalue problem~\cite{footnoteSone}.
Here, $H$ ($S$) is the Hamiltonian (overlap) matrix, $\psi$ is the eigenvector, and $\bm{k}$ is the wave number vector.
We allow the matrices $H$ and $S$ may depend on the eigenvalue $\omega$~\cite{footnote1,footnote2}.

Now, we discuss BEC of nonlinear eigenvalue problems. As a first step, we consider the matrix pencil $P(\omega,\bm{k})$~\cite{Ikramov_MatPencil(1993)},
\begin{equation}
P(\omega,\bm{k}) = H(\omega,\bm{k})- \omega S(\omega,\bm{k}),
\end{equation}
where the solution of $P(\omega,\bm{k})\psi=0$ is equivalent to Eq.~(1). 
In order to discuss the BEC of Eq.~(1), we introduce the auxiliary eigenvalue $\lambda$ and analyze its eigenvalue problems~\cite{footnote3,footnoteLambda},
\begin{equation}
P(\omega,\bm{k})\psi = \lambda\psi.
\end{equation}
Here, $\lambda$ is auxiliary and does not have physical meaning except $\lambda=0$.
The physical eigenvalue $\omega$ is a free parameter.

Next, we analyze Eq.~(3) and discuss the BEC in the auxiliary bands of $\lambda$.
Here, we need to assume $\lambda=0$ exists in the band gap of auxiliary eigenvalues.
We note that the eigenstates only on $\lambda=0$ emerge in physics because auxiliary eigenvectors with finite $\lambda$ do not satisfy Eq.~(1).
When the bulk bands of $\lambda$ are topological, they possess gapless edge states under the open boundary condition.
Since the edge states are gapless, those edge states cross $\lambda=0$ inevitably (see Fig.~1).
Therefore, those edge states can be expected to emerge in physics.

Here, a comment is in order about the range of validity of the above discussion.
The above discussion is valid when nonlinearity is weak in the vicinity of $\omega$ of interest (i.e., auxiliary eigenvalues are monotonic with respect to $\omega$).
This is because the band indices of the bands of $\lambda$ and the bands of $\omega$ correspond one-to-one when the nonlinearity is weak.
In this case, the gapless (gapped) nature of edge states of $\lambda$ is inherited to the bands of $\omega$.
In contrast, when the nonlinearity is strong, band indices of the bands of $\lambda$ and the bands of $\omega$ do not correspond in general. 
In this case, the edge states of $\omega$ can be gapless even if the edge states of $\lambda$ are gapped.
Thus, these gapless edge states cannot be characterized by the topological number, which is calculated by the eigenstates of $\lambda$ (for details, see Sec.~I of Supplemental Material~\cite{supple1}).
In addition, Eq.~(1) can possess complex eigenvalues even when the matrices $H$ and $S$ are Hermitian under the strong nonlinearity.
While the above cases are intriguing, strong nonlinearity induces additional complexities, which requires different approaches.
Therefore, in the following discussion, we focus on the case where the nonlinearity is weak, i.e., the band indices of the bands of $\lambda$ and the bands of $\omega$ correspond, and $\lambda$ is real.

%%%%%%%%%%%%%%%%%%%%%%%%%%%%%%%%%%%%%%%%%%%%%%%%%%%%%%%%%%%%%%%%%%%%%%%%%
%%%%%%%%%%%%%%%%          GEVP with Hermitian matrices      %%%%%%%%%%%%% %%%%%%%%%%%%%%%%%%%%%%%%%%%%%%%%%%%%%%%%%%%%%%%%%%%%%%%%%%%%%%%%%%%%%%%%% 　\\ %%%%%%%%%%%%%%%%%%%%%%%%%%%%%%%%%%% %%%%%%%  paragraph1  %%%%%%%%%%%%%%
\begin{figure}[t]
   \includegraphics[width=9cm]{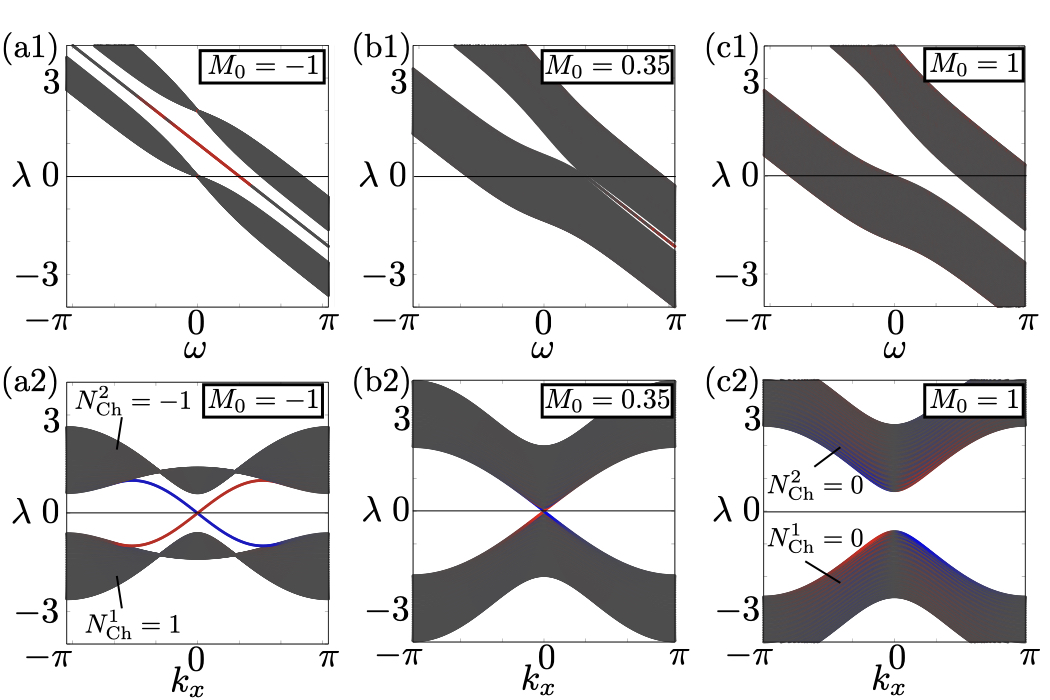}
 \caption{
 Band structure of the auxiliary eigenvalue $\lambda$ of the two-dimensional model.
 Edge (bulk) states are plotted in red (gray).
 (a1)-(c1) [(a2)-(c2)]: Plot of $\lambda$ for each $\omega$ [$k_x$] with $M_0=-1$, $M_0=0.35$, and $M_0=1$ respectively.
 The parameter $k_x$ ($\omega$) is fixed in $k_x=0$ ($\omega=1$).
Parameter $k_x$ ($\omega$) is choosen in $k_x=0$ ($\omega$=1).
Edge states emerge when $M_0<0.35$.
}
\label{label:Fig_lam-w}
\end{figure}

\begin{figure}[t]
   \includegraphics[width=9cm]{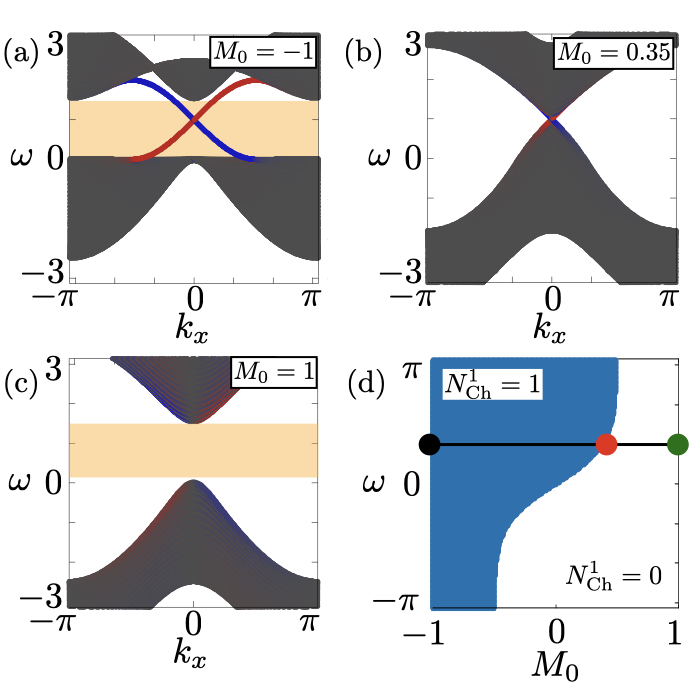}
 \caption{
 Band structure of $\omega$ of the two-dimensional model.
 Edge (bulk) states are plotted in red (gray).
 (a)-(c): Plot of $\omega$ for each $k_x$ and the Chern number of the bands of $\lambda$ with $M_0=-1$, $M_0=0.35$, and $M_0=1$ respectively.
The band gap is fully opening in the orange-colored regions.
(d): Plot of the Chern number calculated from the eigenstates of $\lambda_1$ for each $M_0$ and $\omega$.
The Chern number takes a non-zero (zero) value in the blue-colored region.
The black line represents $\omega=1$.
The band structures in Figs.~3(a), 3(b), and 3(c) emerge on the black, red, and green dots.
The region where the Chern number takes non-zero values corresponds to the region where the edge states emerge. 
 }
\label{label:Fig_lam-w}
\end{figure}

\textit{Nonlinear Chern insulator and gapless edge states.}---Here, we explore the nonlinear BEC in a two-dimensional model.
We examine the relationship between the bulk topology of the auxiliary bands of $\lambda$ and the gapless edge states of $\omega$ in a two-dimensional system described by a nonlinear eigenvalue problem [Eq.~(1)].

Here, we analyze the following two-dimensional model with $\omega$-dependent terms,
\begin{equation}
H(\bm{k})=
\begin{pmatrix}
E+M_\mathrm{H}(\bm{k})&\mathrm{sin}(k_x)-i\mathrm{sin}(k_y)\\
\mathrm{sin}(k_x)+i\mathrm{sin}(k_y)&E-M_\mathrm{H}(\bm{k})
\end{pmatrix},
\end{equation}
\begin{equation}
S(\omega)=
\begin{pmatrix}
1-M_\mathrm{S}(\omega)&0\\
0&1+M_\mathrm{S}(\omega)
\end{pmatrix},
\end{equation}
with $E=1$, $M_{\mathrm{H}}(\bm{k})=M_0+\sum_{i=x,y}[1-\mathrm{cos}(k_i)]$, and $M_{\mathrm{S}}(\omega)=M_1\mathrm{tanh}(\omega)/\omega$.
Here, $M_1$ is fixed to $0.5$.
We denote this system by a nonlinear Chern insulator (for the case of the chiral symmetric system, see Sec.~II of Supplemental Material [60]).
From these matrices, matrix $P$ is given by,
\begin{equation}
P(\omega,\bm{k})=
\begin{pmatrix}
E_{\mathrm{P}}(\omega)+M_{\mathrm{P}}(\omega,\bm{k})&\mathrm{sin}(k_x)-i\mathrm{sin}(k_y)\\
\mathrm{sin}(k_x)+i\mathrm{sin}(k_y)&E_{\mathrm{P}}(\omega)-M_{\mathrm{P}}(\omega,\bm{k})
\end{pmatrix},
\end{equation}
with $E_{\mathrm{P}}(\omega)=E-\omega$ and $M_{\mathrm{P}}(\omega,\bm{k})=M_{\mathrm{H}}(\bm{k})+\omega M_{\mathrm{S}}(\omega)$.

Here, let us analyze the auxiliary band structure of $\lambda$ by solving the eigenvalue problem $P(\omega,\bm{k})\psi=\lambda\psi$.
The band structure of $\lambda$ is plotted in Fig.~2.
For the calculations of Fig.~2, we consider the open (periodic) boundary conditions in the $y$-direction ($x$-direction).
Figures~2(a1)-2(c1) [2(a2)-2(c2)] displays auxiliary eigenvalues $\lambda$ for each $\omega$ ($k_x$) with $k_x=0$ ($\omega=1$).
Bulk (edge) states are plotted in gray (red).
When $M_0=-1$, edge states emerge [see Figs.~ 2(a1) and 2(a2)].
Since these edge states are gapless in the momentum space, the edge states cross $\lambda=0$ inevitably. 
This result suggests the existence of physical edge states inherited from the auxiliary bands.
In Figs.~2(b1) and 2(b2), the band gap closes at $M_0=0.35$. 
The point of $\lambda(\omega=1,k_x=0)$ becomes the Dirac point.
When $M_0=1$, the band gap reopens, and the edge states vanish.
From these results, we can conclude that the topological phase transition occurs at $M_0 = 0.35$. Therefore, when $M_0 < 0.35$, we can expect the presence of physical edge states since the edge states of auxiliary bands cross $\lambda=0$ inevitably.

In order to characterize the gapless edge states in Fig.~2, let us consider the topological number of the auxiliary bands of $\lambda$.
For our two-dimensional model, the Chern number can be used as a topological number.
The Chern number of the band index $n$ is defined by,
\begin{equation}
N_{\mathrm{Ch}}^{n}(\omega)=\frac{1}{2\pi}\int_{1BZ} dk_xdk_y\bm{\nabla}_k\times
\bm{A}_n(\omega,\bm{k}),
\end{equation}
\begin{equation}
\bm{A}_n(\omega,\bm{k})=\langle\psi_{n,k}(\omega)|\bm{\nabla}_k\psi_{n,k}(\omega)\rangle,
\end{equation}
where $|\psi_{n,\bm{k}}(\omega)\rangle$ is an auxiliary eigenstate with band index $n$ and momentum $\bm{k}$.
The integration is conducted in the first Brillouin zone of the momentum space.
In our model, Berry connection $\bm{A}_n(\omega,\bm{k})$ depend on $\omega$.
Thus, the Chern number becomes the function of $\omega$.

In our model, the Chern number takes non-zero values when $M_0<0.35$.
In Fig.~2(a2), the Chern number of the lower (upper) band is $N_{\mathrm{Ch}}^{1}(\omega_{\mathrm{R}})=1$ [$N_{\mathrm{Ch}}^{2}(\omega_{\mathrm{R}})=-1$] while $N_{\mathrm{Ch}}^{1}(\omega_{\mathrm{R}})=N_{\mathrm{Ch}}^{2}(\omega_{\mathrm{R}})=0$ in Fig.~2(c2) with $\omega_{\mathrm{R}}=\omega=1$.
Therefore, the Chern number calculated by the eigenstates of $\lambda$ corresponds to the number of edge states of the auxiliary bands of $\lambda$.

Here, let us investigate the nonlinear BEC between the above Chern number and the edge states of $\omega$.
The physical band structures of $\omega$ are plotted in Fig.~3 by extracting the data of $\lambda=0$.
In Figs.~3(a)-3(c), the bands of $\omega$ are plotted for each $k_x$ with $M_0=-1$, $M_0=0.35$, and $M_0=1$ respectively.
Bulk (edge) states are plotted in gray (red or blue).
For the calculation of these figures, the open (periodic) boundary condition is imposed in the $y$- ($x$-) direction. 
The region where the band gap of $\omega$ remains open across all $k_x$ is colored with orange.
In our model, the gapless edge states of $\omega$ emerge when $M_0=-1$ [see Fig.~3(a)].
The edge states are inherited from the edge states of $\lambda$ plotted in Fig.~2(a2).
The band structure of $\omega$ become gapless when $M_0=0.35$ [see Fig.~3(b)].
This gapless point corresponds to the gap-closing point in Fig.~2(b2), which elucidates a nonlinear topological phase transition.
When $M_0=1$, the edge states vanish  [see Fig.~3(c)].

Figure~3(d) is the plot of the Chern number of the lower band obtained from the eigenstates of $\lambda$ for each $M_0$ and $\omega$.
The Chern number takes a non-zero value in the blue-colored region.
The black line represents $\omega=1$.
The band structures plotted Figs.~3(a), 3(b), and 3(c) emerge on the black, red, and green dots in Fig.~3(d).
Importantly, correspondence exists between the regions where the Chern number is non-zero and the regions where edge states emerge.
The above results demonstrate that the nonlinear BEC holds for nonlinear Chern insulator.

Here, we note that when discussing the nonlinear BEC from the phase diagram plotted in Fig.~3(d), $\omega_{\mathrm{R}}$ is properly chosen so that it is inside of the band gap of $\omega$.
Taking $\omega_R$ outside the band gap can lead to a mismatch between the Chern number and the edge states.

\begin{figure}[t]
   \includegraphics[width=8cm]{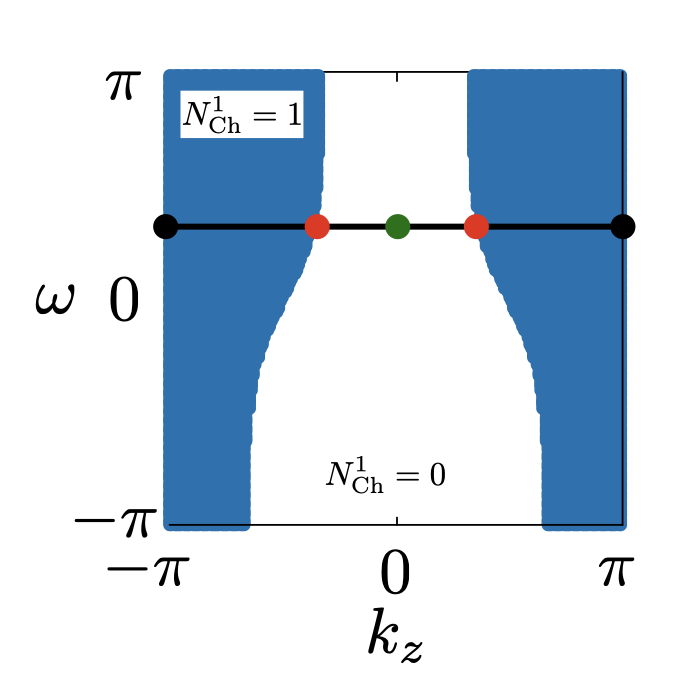}
 \caption{
 Plot of the Chern number calculated by the eigenstates of $\lambda$ for each $k_z$ and $\omega$.
 The Chern number takes 1 (0) in the blue (white) region.
 The black line represents $\omega=1$.
 Band structures of Figs.~3(a), 3(b), and 3(c) emerge on the black, red, and green dots.  
The boundaries between the blue and the white regions correspond to the Weyl nodes.
}
\end{figure}

\textit{Nonlinear topological semimetal and Fermi arc surface states.}---Next, let us consider the nonlinear BEC in a three-dimensional model. 
We demonstrate the existence of the nonlinear BEC between the bulk topology of the auxiliary bands and the Fermi-arc surface states of the nonlinear Weyl semimetals (see below) by analyzing a three-dimensional model including $\omega$-dependent term.

Here, we analyze Eqs.~(4)-(6) with $M_0=\mathrm{cos}(k_z)$.
In this model, the gapless band structure plotted in Fig.~3(b) emerges at $\mathrm{cos}(k_z)=-0.35$.
These gapless points of $\omega$ correspond to the Weyl points, which form a line structure in the four-dimensional parameter space, composed of $\omega$, $k_x$, $k_y$ and $k_z$.
The band structure in Fig.~3(a) [(b)] also emerges at $k_z=\pi$ [$k_z=0$].
In this manner, we define a system described by the nonlinear eigenvalue problems that possess pairs of Weyl points as nonlinear Weyl semimetals.

Figure~4 is the plot of the Chern number of the lower band obtained from the eigenstates of $\lambda$ for each $k_z$ and $\omega$.
The blue (white) region represents where the Chern number takes $1$ ($0$).
The black line represents $\omega=1$, and colored dots indicate the points where the band structures plotted in Fig.~3 emerge.
On the black, red, and green points, The band structure plotted in Fig.~3(a), 3(b), and 3(c) emerges.
In Fig. 4, the boundaries between the blue and white colored region correspond to the Weyl nodes.
In our model, Weyl nodes form line structures in the $k_z$-$\omega$ space.
However, in general, these lines of Weyl points emerge in the four-dimensional parameter space composed of $k_x$, $k_y$, $k_z$, and $\omega$.
Notably, as is the case of the two-dimensional model, the correspondence exists between the region where the Chern number takes a non-zero value and the region where edge states emerge.
This result shows the applicability of nonlinear BEC even in the topological semimetals.

\textit{Summary.}---
In this letter, we have established the BEC for nonlinear systems of eigenvalues. 
We have introduced an auxiliary eigenvalue and showed the emergence of BEC between the bulk topology of auxiliary eigenstates and physical edge states when the nonlinearity is weak. 
Applying our argument to two-dimensional insulators and three-dimensional semimetals, we have demonstrated the emergence of edge states induced by the nonlinear BEC.
Our results show BEC remains valid even beyond the linear systems.
Our nonlinear BEC is applicable to the model with more complicated $\omega$-dependence as long as the nonlinearity is weak (i.e., auxiliary eigenvalues are monotonic with respect to $\omega$). 
Our approach with the auxiliary eigenvalue is considered to be extended to systems in other symmetry classes and dimensions.
These extensions and applications to more physical setups are left as future works.

%%%%%%%%%%%%%%%%%%%%%%%%%%%%%%%%%%%%%%%%%%%%%%%%%%%%%%%%%%%%%%%%%%%%%%%%%
%%%%%%%%%%%%%%%%          Acknowledgement                    %%%%%%%%%%%%
%%%%%%%%%%%%%%%%%%%%%%%%%%%%%%%%%%%%%%%%%%%%%%%%%%%%%%%%%%%%%%%%%%%%%%%%%

\textit{Acknowledgement.}---This work is supported by MEXT-JSPS Grant-in-Aid for Transformative Research Areas (A) ``Extreme Universe": Grant No.~JP22H05247. This work is also supported by JST-CREST Grant No.~JPMJCR19T1, JST-SPRING Grant No.~JPMJSP2124, and JSPS KAKENHI Grant No.~JP21K13850, JP23H01091.
TY is grateful to the long term workshop YITP-T-23-01 held at YITP, Kyoto University.

%\label{sec:noncontractible}
\bibliography{main} %hoge.bibから拡張子を外した名前

\begin{thebibliography}{10}
\bibitem{footnote1}
{ \red{If $H$ and $S$ are independent from $\omega$, Eq.~(1) is reduced to
  generalized eigenvalue problem. If $H$ is independent from $\omega$ and $S$
  is the identity matrix, Eq.~(1) is reduced to the ordinary eigenvalue
  problem.}}
\bibitem{footnoteLambda}
{ \red{We note that how to introduce $\lambda$ is not uniquely determined.
  \blue{For example, it is possible to introduce a matrix $Q$ so that Eq.~(3)
  becomes a generalized eigenvalue problem $P\psi=\lambda Q\psi$.}}}
\end{thebibliography}
%\bibliographystyle{ieeetr} %
%\bibliographystyle{savethrees} %
%\bibliography{empty} %
%\bibliographystyle{abbrv} %参考文献出力スタイル
\bibliographystyle{h-physrev5}

\end{document}

%%%%%%%%%%%%%%%%%%%%%%%%%%%%%%%%%%%%%%%%%%%%%%%%%%%%%%%%%%%%%%%%%%%%%%%
\title{
  Supplemental Materials for 
``Bulk-\blue{edge} correspondence for nonlinear eigenvalue problems"
}

\author{Takuma Isobe}
\affiliation{Graduate School of Pure and Applied Sciences, University of Tsukuba, Tsukuba, Ibaraki 305-8571, Japan}
\author{Tsuneya Yoshida}
\affiliation{Department of Physics, Kyoto University, Kyoto 606-8502, Japan}
\author{Yasuhiro Hatsugai}
\affiliation{Department of Physics, University of Tsukuba, Tsukuba, Ibaraki 305-8571, Japan}
%\author{Takuma Isobe$^1$}
%\author{Tsuneya Yoshida$^2$}
%\author{Yasuhiro Hatsugai$^3$}
%\affiliation{                                                                   
%$^1$Graduate School of Pure and Applied Sciences, University of Tsukuba, Tsukuba, Ibaraki 305-8571, Japan\\
%$^2$Department of Physics, Kyoto University, Kyoto 606-8502, Japan\\
%$^3$Department of Physics, University of Tsukuba, Tsukuba, Ibaraki 305-8571, Japan
%}
%%%%%%%%%%%%%%%%%%%%
%%%%%%%%%%%%%%%%%%%%%%%%%%%%%%%%%%%%%%%%%%%%%%%%%%%%%%%%%%%%%%%%%%%%%%%
\date{\today}
%%%%%%%%%%%%%%%%%%%%%%%%%%%%%%%%%%%%%%%%%%%%%%%%%%%%%%%%%%%%%%%%%%%%%%%
%\begin{abstract}
%\end{abstract}
%%%%%%%%%%%%%%%%%%%%%%%%%%%%%%%%%%%%%%%%%%%%%%%%%%%%%%%%%%%%%%%%%%%%%%%
%\pacs{
%***
%} 
%%%%%%%%%%%%%%%%%%%%%%%%%%%%%%%%%%%%%%%%%%%%%%%%%%%%%%%%%%%%%%%%%%%%%%%
%%%%%%%%%%%%%%%%%%%%%%%%%%%%%%%%%%%%%%%%%%%%%%%%%%%%%%%%%%%%%%%%%%%%%%%
\maketitle
%%%%%%%%%%%%%%%%%%%%%%%%%%%%%%%%%%%%%%%%%%%%%%%%%%%%%%%%%%%%%%%%%%%%%%%

%%%%%%%%%%%%%%%%%%%%%%%%%%%%%%%%%%%%%%%%%%%%%%%%%%%%%%%%%%%%%%%%%%%%%%%%%
%%%%%%%%%%%%%%%%          Appendix1                          %%%%%%%%%%%%
%%%%%%%%%%%%%%%%%%%%%%%%%%%%%%%%%%%%%%%%%%%%%%%%%%%%%%%%%%%%%%%%%%%%%%%%%
%%%%%%%%%%%%
\section{Strength of nonlinearity for nonlinear BEC}

In this section, we discuss the \red{relation} between the strength of nonlinearity of $\omega$ and the nonlinear \blue{BEC}. 
In order to discuss the nonlinear \blue{BEC}, we have made the assumption that the nonlinearity of $\omega$ is weak near the $\omega$ of interest in the main text. When the nonlinearity of $\omega$ is weak, the band indices of $\lambda$ correspond to the band indices of $\omega$ [see Fig.~1(a)].  
Consequently, the presence of gapless \blue{edge} states between $\lambda_i$ and $\lambda_{i+1}$ results in the gapless \blue{edge} states within $\omega_i$ and $\omega_{i+1}$. 
This is due to the fact that the gapless \blue{edge} states of $\lambda$ intersect the $\lambda=0$ for any $\omega$ within the ranges of $\omega_i$ and $\omega_{i+1}$.
When the \blue{edge} states of $\lambda$ are gapped (edge states in the gap do not connect the upper and lower bands), there exist $\omega$ that do not cross $\lambda=0$.
Therefore, the \blue{edge} states of $\omega$ also become gapped. 

On the contrary, when the nonlinearity of $\omega$ is strong, the band indices of $\lambda$ do not correspond with the band indices of $\omega$ [see Fig.~1(b)]. 
In such cases, even if the \blue{edge} states of $\lambda$ are gapped, the \blue{edge} states of $\omega$ can still be gapless. 
As a result, when this situation arises, there exists a case that the bands of $\omega$ are gapples even if the bands of $\lambda$ are gappless.

\begin{figure*}[h]
   \includegraphics[width=18cm]{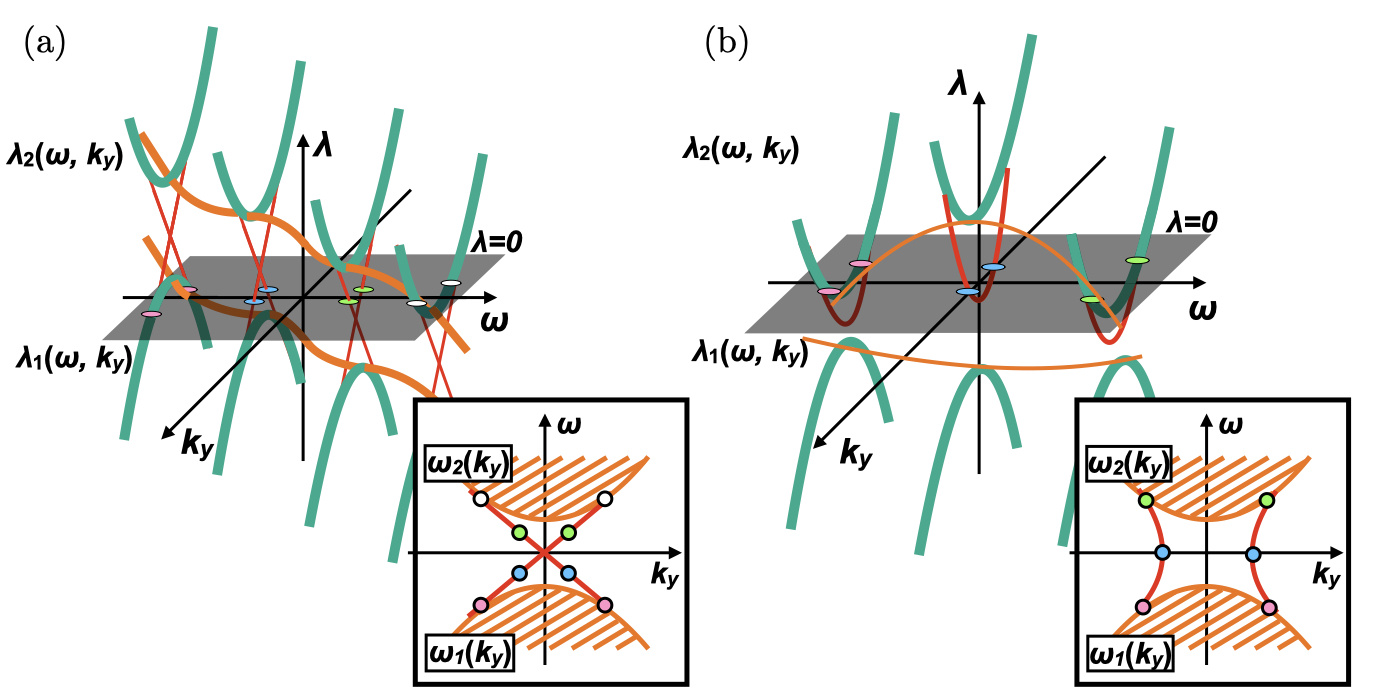}
 \caption{
Sketch of the auxiliary and physical band structures.
\blue{Edge} states are illustrated in red.
The band structure of $\omega$ is sketched in insets.
(a): Band structures when $\lambda$ is monotonic for each $\omega$.
The band indices of $\lambda$ correspond to the band indices of $\omega$.
(b): Band structures when $\lambda$ is not monotonic for each $\omega$.
The band indices of $\lambda$ do not correspond to the band indices of $\omega$.
Although the \blue{edge} states of $\lambda$ are gapped, \blue{edge} states of $\omega$ are gapless.
}
\label{label:Fig_model}
\end{figure*}

\section{1D model and topological zero modes}

%%%%%%%%%%%%%%%%%%%%%%%%%%%%%%%%%%%
%%%%%%%%%%   1   %%%%%%%%%%
In this section, we demonstrate the nonlinear \blue{BEC} between the bulk topology of the auxiliary eigenvalue $\lambda$ and physical \blue{edge} states. 
We achieve this by analyzing a one-dimensional model that contains the nonlinearity of the physical eigenvalue $\omega$. 
One-dimensional systems can exhibit topological zero modes. Therefore, if the auxiliary bands of $\lambda$ possess topological zero modes, these modes can manifest as physical behaviors.

%%%%%%%%%%   2   %%%%%%%%%%
Here, we analyze one-dimensional model with $\omega$ dependent term $M_{\mathrm{S}}(\omega)$,
\begin{equation}
H(k)=
\begin{pmatrix}
0& \mathrm{sin}(k)-iM_\mathrm{H}(k)\\
\mathrm{sin}(k)+iM_\mathrm{H}(k)&0
\end{pmatrix},
\end{equation}
\begin{equation}
S(\omega)=
\begin{pmatrix}
1-M_\mathrm{S}(\omega)&0\\
0&1+M_\mathrm{S}(\omega)
\end{pmatrix},
\end{equation}
where $\omega$ and $k$ represent frequency and wavenumber, respectively.
We consider the cases $M_{\mathrm{H}}(k)=M_0+[1-\mathrm{cos}(k)]$ and $M_{\mathrm{S}}(k)=M_1\mathrm{tanh}(\omega)/\omega$.
We note that $\lambda$ decreases monotonically with respect to $\omega$ when $M_1\leqq$1.
Thus, we fix $M_1=0.5$.
Because the matrix $S$ depends on $\omega$, the eigenvalues cannot be obtained simply by diagonalization.
We analyze this nonlinear eigenvalue problem using the auxiliary eigenvalue $\lambda$.
From these matrices, matrix $P$ is composed as follows,
\begin{equation}
P(\omega,k)=
\begin{pmatrix}
-\omega+M_{\mathrm{S}}(\omega)& \mathrm{sin}(k)-iM_\mathrm{H}(k)\\
  \mathrm{sin}(k)+iM_\mathrm{H}(k)&-\omega-M_{\mathrm{S}}(\omega)&\\
\end{pmatrix}.
\end{equation}

%%%%%%%%%%   3   %%%%%%%%%%
Let us analyze the one-dimensional model described by $P(\omega,k)\psi=\lambda\psi$.
In Figs. 2(a)-2(c), the auxiliary eigenvalue $\lambda$ is plotted for each $\omega$ under the open boundary condition.
Bulk (\blue{edge}) states are plotted in gray (red).
When $M_0=-0.8$, the \blue{edge} states emerge [see Fig.~2(a)].
The \blue{edge} states degenerate at $\omega=0$ because $M_{\mathrm{S}}(\omega)$ becomes zero at $\omega=0$.
The band gap closes at $M_0=0$ [see Fig.~2(b)], and the \blue{edge} states vanish when $M_0=0.8$ [see Fig.~2(c)].
From these figures, we can conclude that topological phase transition occurs at $M_0=0$. 
Therefore, when $M_0<0$, we can expect the presence of physical \blue{edge} states at $\omega=0$ since the \blue{edge} states cross $\lambda=0$.

In order to characterize the \blue{edge} modes in Fig.~2(a), let us discuss the bulk topology of auxiliary bands of $\lambda$.
In this model, matrix $P(\omega,k)$ satisfy the chiral symmetry,
\begin{equation}
\Gamma^{-1}P(\omega_0,k)\Gamma=-P(\omega_0,k),
\end{equation}
with $\omega=\omega_0=0$.
The chiral operator is denoted by $\Gamma$.
Notably, chiral symmetry guarantees the emergence of the \blue{edge} states of $\lambda$ at $\lambda(\omega_0, k)=0$.
Thus, the topological zero modes protected by the chiral symmetry of $P$ emerge as the physical \blue{edge} modes.
We note that the chiral symmetry of the matrix $P$ does not have any physical meaning. 
The symmetry purely imposes a mathematical constraint on $P$ and $\lambda$.

%%%%%%%%%%   4   %%%%%%%%%%
Here, let us verify the correspondence between the bulk topology of the auxiliary bands of $\lambda$ and the physical \blue{edge} states of $\omega$.
Since our model satisfies the chiral symmetry at $\omega=\omega_0$, the winding number can be defined as follows,
\begin{equation}
\nu = \frac{i}{2\pi}\int_{-\pi}^{\pi} dk \partial_k \mathrm{log} P_{12},
\end{equation}
where $P_{12}$ is $1,2$ component of $P(\omega,\bm{k})$.
We employ the winding number to discuss the bulk topology of the band structure of $\lambda$.

In Fig.~2(d), the frequency $\omega$ and the winding number of $P_{12}$ are plotted for each $M_0$.
When $M_0<0$, since the spectrum of $P_{12}$ wind around the origin, the winding number takes a non-zero value $\nu(\omega_0)=1$.
Corresponding to the non-zero winding number, the physical \blue{edge} modes of $\omega$ emerge at $\omega=0$.
Contrarily, since the spectrum of $P_{12}$ does not wind the origin when $M_0>0$, the winding number becomes zero, and the \blue{edge} state vanishes.
This result implies that the bulk topology of the band of $\lambda$ corresponds to the physical \blue{edge} state of $\omega$.
%%%%%%%%%%   5   %%%%%%%%%%

%%%%%%%%%%   6   %%%%%%%%%%
Based on the above discussion, we can conclude that a \blue{BEC} exists between the topological number of the auxiliary bands of $\lambda$ and the physical \blue{edge} states of $\omega$. 
%In this section, we have focused on the indirect \blue{BEC} in the one-dimensional case. 

%%%%%%%%%%%%%%%%%%%%%%%%%%%%%%%%%%%
%%%%%%%   figure 3   %%%%%%%%%%%%%%
%%%%%%%%%%%%%%%%%%%%%%%%%%%%%%%%%%%
\begin{figure}
   \includegraphics[width=9cm]{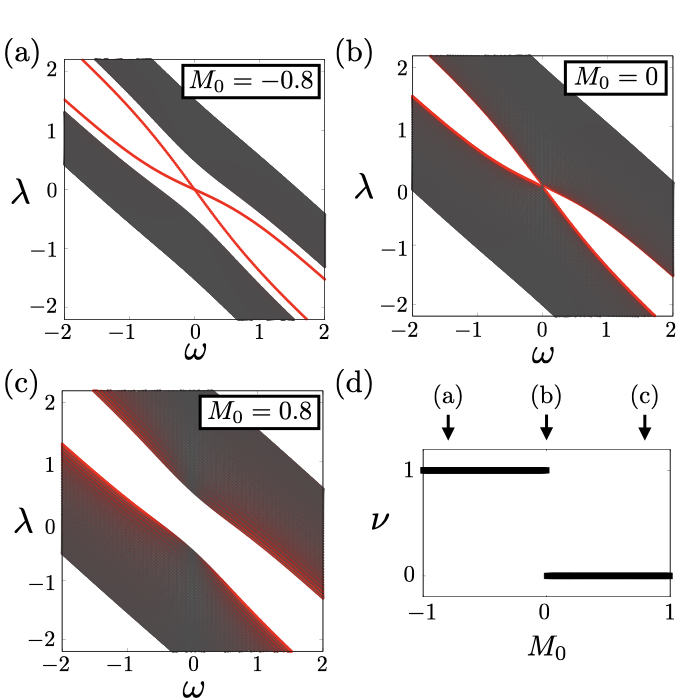}
 \caption{
Band structure of the one-dimensional model.
\blue{Edge} (bulk) states are plotted in red (gray). 
(a)-(c): Band structures of $\lambda$ for each $\omega$ with $M_0=-0.8$, $M_0=0$, and $M_0=0.8$ respectively.
\blue{Edge} states emerge when $M_0<0$.
(d): Plot of $\omega$ for each $M_0$ and the winding number $\nu$ with $\omega=\omega_0=0$.
The winding number takes a non-zero value where the region that the \blue{edge} states emerge.}
\label{label:Fig_model}
\end{figure}

%%%%%%%%%%%%